\documentclass[amsmath,amssymb,aip,jcp,citeautoscript,superscriptaddress,twocolumn,10pt]{revtex4-1}

\usepackage{graphicx}
\usepackage{dcolumn}
\usepackage{bm}

\usepackage{mathrsfs}
\usepackage{color}

\usepackage[normalem]{ulem}

\graphicspath{{figures/}}
\begin{document}

\title{
Dissipative time-dependent quantum transport theory:
quantum interference and phonon induced decoherence dynamics
}

\author{Yu Zhang}
\email{zhy@yangtze.hku.hk}
\altaffiliation[Present institute: ]{Center of Bio-inspired Energy Science, Northwestern University, Evanston, IL, USA.}
\affiliation{
Department of Chemistry, The University of Hong Kong, Pokfulam Road, Hong Kong, China
}

\author{ChiYung Yam}
\affiliation{
Beijing Computational Science Research Center, Beijing 100084, China
}
\affiliation{
Department of Chemistry, The University of Hong Kong, Pokfulam Road, Hong Kong, China
}

\author{GuanHua Chen}
\email{ghc@everest.hku.hk}
\affiliation{
 Department of Chemistry, The University of Hong Kong, Pokfulam Road, Hong Kong, China
}

\date{\today}

\begin{abstract}
A time-dependent inelastic electron transport theory for
strong electron-phonon interaction is established
via the equations of motion method combined with the small polaron
transformation. In this work, the dissipation via electron-phonon
coupling is taken into account in the strong coupling regime, which
validates the small polaron transformation. The corresponding equations of
motion are developed, which are used to study the
quantum interference effect and phonon-induced decoherence dynamics in
molecular junctions. Numerical studies show clearly
quantum interference effect of the transport electrons through
two quasi-degenerate states with different coupling to the leads.
We also found that the quantum interference can be suppressed by the electron-phonon
interaction where the phase coherence is destroyed by phonon scattering.
This indicates the importance of electron-phonon
interaction in systems with prominent quantum interference effect.
\end{abstract}


\maketitle

\section{Introduction}
Interplay between inelastic scattering and coherence in quantum
transport is closely related to the performance of molecular 
electronics. In the presence of phonons, electrons have the probability
of being scattered off inelastically by phonons.
Inelastic scattering of transport electrons and energy dissipation 
play a vital role in device characteristics, working performance 
and stability.
Effects of electron-phonon interaction in 
the single molecule junction have attracted a lot of attention
both experimentally and theoretically~\cite{PhysRevLett.108.146602,
PhysRevLett.88.216803,magnus2006jpcs, PhysRevB.72.201101,magnus2008prl,
viljas2005prb,michael2007jpcm,PhysRevB.86.161404,PhysRevB.86.155411,
C2CP41017F,RevModPhys.83.131}.
Even at zero temperature, the vibrational
motions of molecules are essentially frozen, phonon can be excited by the
electronic current. The energy exchange between 
electrons and phonons
is directly responsible for the local heating or cooling~\cite{PhysRevB.81.115438,PhysRevLett.93.256601,
mceniry2007jpcm,Eunan195304,PhysRevB.80.115427}.
When the electron-phonon coupling strength is strong,
a vibronic state (polaron) may be formed when the electron resides
in the junction for relatively long time. 
The formation of polaron is determined by the
detailed balance between transport electronic energy and vibrational
relaxation.

To understand the nature of dissipative transport, theoretical
methods including single-particle and many-particle approaches
were developed. Many-particle approaches include quantum 
master equation, path-integral method. However, these
approaches are computationally expensive since the 
dimension increases exponentially with the system size,
which limit their applications to large systems. 
Instead, we recently established a dissipative time-dependent
quantum transport theory~\cite{zhang:164121} based on 
single-particle picture. This theory is an extension of the newly 
proposed time-dependent density functional theory for open
quantum system (TDDFT-OS) combined with nonequilibrium Green's
function (NEGF) method, termed
TDDFT-OS-NEGF~\cite{PhysRevB.75.195127,C1CP20777F,yijing2008,
zheng:114101,C2NR32343E, yam2008,
PhysRevB.87.085110,xie:044113,tian:204114,tddft2013},
which propagates the equations of motion (EOMs) for signle-electron
density matrix~\cite{Wang2007,Liang2002,Liang2000}.
The dissipation via phonon is taken into account by introducing
a self-energy for the electron-phonon interaction in addition
to the self-energies induced by the electrodes\cite{zhang:164121}.
Due to its single-particle nature, the dissipative 
time-dependent quantum transport theory is efficient for the
investigation of the transient dynamics of electron 
transport with electron-phonon interaction in large systems
and can be readily extended to time-dependent density 
functional theory. In practice, the wide-band limit (WBL) approximation is 
usually applied to further reduce the computational cost and the resulting TDDFT-OS-NEGF-WBL has been applied
successfully to study transient electron dynamics in molecular electronic devices~\cite{zhang:164121,PhysRevB.87.085110}.

However, the dissipative time-dependent quantum transport theory 
proposed in Ref.~\onlinecite{zhang:164121} is based on the lowest
order expansion 
with respect to electron-phonon coupling, 
where its applications are limited to the weak electron-phonon coupling 
regime.
In the strong electron-phonon coupling regime, polaron transformation
is usually adopted, which has been applied to one-level model coupled
with one phonon mode for both steady state and transient dynamic 
properties of junctions \cite{PhysRevB.66.075303,PhysRevB.71.165324,
nitzan2006prb,PhysRevB.67.165326,goker2011jpcm}.
This has also been extended to study the steady state properties of 
multi-level model~\cite{dahnovsky:234111,*dahnovsky:014104,Alan2012arxiv},
while those studies are limited to the steady state.
A time-dependent method is desirable for the 
investigation of quantum dynamics of dissipative systems with
strong electron-phonon coupling. In this work,
a dissipative time-dependent quantum transport theory for strong
electron-phonon coupling is established
by combining TDDFT-OS-NEGF-WBL and polaron transformation.

The method developed in this work is applied to investigate
the quantum interference effects and phonon-induced decoherence
dynamics in molecular junctions, which is a fundamental 
quantum-mechanical effect and has received great attention 
recently~\cite{PhysRevLett.109.056801,PhysRevLett.107.046802,Guedon2012,
PhysRevLett.109.186801,nl0608442,kesanhuang2008,jz5007143}. 
The quantum interference effects have
been observed in a closely related field of the electron transport
through quantum dots which are set up as Aharonov-Bohm interferometers
\cite{nature03899,buks1998nature,schuster1997nature,PhysRevLett.87.256802}.
A great deal of both theoretical and experimental efforts
have been made to study the quantum interference effects in
the molecular junctions due to its fundamental importance
and practical applications such as quantum interference transistor
\cite{PhysRevLett.109.056801,PhysRevLett.107.046802,Guedon2012,
PhysRevLett.109.186801,nl0608442,kesanhuang2008,jz5007143}. After the electron 
injection
from the leads to the system, the electrons undergo a transient
nonequilibrium transport process before the quantum interference pattern
is formed. While intensive studies have been carried out on the exploring of
to study the
steady state quantum interference effect, the dynamics of electron transport
in a quantum interference system and phonon induced docoherence process
remain largely unexplored. The dissipative time-dependent
quantum transport theory developed in this work is thus well-suited
for these purposes.

The article is organized as follows.
Sec.~\ref{sec:method} introduces the dissipative time-dependent quantum
transport theory with electron-phonon interaction in strong coupling
regime, starting from a single-electron Hamiltonian. 
The method presented
in Sec.~\ref{sec:method} is then applied to study the 
quantum interference effect and phonon-induced decoherence dynamics in molecular junctions.
Numerical studies and related discussions are given in
Sec.~\ref{sec:results}. Finally, we summarize this work in Sec.~\ref{sec:conclusion}.

\section{Methodology}
\label{sec:method}

\subsection{Model Hamiltonian and Polaron transformation}
The system of interest is a device sandwiched between two leads
and the electrons have the probability of being 
scattered by phonons  when transport through the device. This
transport problem can be modeled by a set of discrete levels localized
in the device region and a continuum of electronic states localized in
each lead. Besides, the vibrational degrees of freedom are described
as harmonic oscillators. Therefore, the corresponding model Hamiltonian
can be written as
\begin{eqnarray}\label{eq:modelhamiltonian1}
H=&&\sum_m \epsilon_m c^\dag_m c_m + \sum_{m\neq n} U_{mn}c^\dag_m c_m c^\dag_n c_n
+\sum_{k,\alpha}\epsilon_{k_\alpha}c^\dag_{k_\alpha} c_{k\alpha}
\nonumber\\
&&+\sum_{m,k,\alpha}[V_{k_\alpha m}c^\dag_{k_\alpha}c_m+\text{H.c.}]+\sum_q
\omega_q a^\dag_q a_q
\nonumber\\
&&+\sum_{q,m}\lambda_{mq} c^\dag_m c_m(a^\dag_q +a_q).
\end{eqnarray}
Where $\epsilon_m$ denotes the energies of electronic states in the device
and $c^\dag_m$ and $c_m$ are the corresponding creation and annihilation operators.
Similarly, $k$th electronic state on the lead $\alpha$ is described
by the energy $\epsilon_{k_\alpha}$, with the creation and annihilation
operators $c^\dag_{k_\alpha}$ and $c_{k_\alpha}$ respectively.
The electron-electron interaction is given by the 
Hubbard-type interaction terms $U$.
The interaction between the electronic states of the device
and lead $\alpha$ is characterized by the coupling strength $V_{mk_\alpha}$.
Due to the coupling to leads, electronic states of the device are
renormalized and are expressed by the self-energy or line-width function.
The line-width function is given by $\Gamma_{\alpha,mn}(\epsilon)=2\pi\sum_{k_\alpha}
V^*_{k_\alpha m}V_{k_\alpha n}\delta(\epsilon-\epsilon_{k_\alpha})$.
If the semi-infinite lead are modeled as a tight-binding chain
with internal hopping parameter $t$, then the line-width function
is obtained as
\begin{equation}\label{eq:linewidth}
\Gamma_{\alpha,mn}(\epsilon)=V_{\alpha,m}V_{\alpha,n}
\frac{\sqrt{4t^2-(\epsilon-\mu_\alpha)^2}}{t^2},
\end{equation}
where $\mu_\alpha$ is the chemical potential of lead $\alpha$.
Similar to $V_{k_\alpha m}$, $V_{\alpha,m}$ is the coupling strength of state $m$
to lead $\alpha$.

The last two terms in the Eq.(\ref{eq:modelhamiltonian1}) are phonon Hamiltonian
and the interaction between electron and phonon.
$a^\dag_q$ ($a_q$) denotes the creation (annihilation) operator of the $q$th phonon
mode with phonon frequency $\omega_q$, the corresponding vibrational
displacement operator is given by $Q_q=a^\dag_q+a_q$. The electron-phonon coupling
constant between phonon mode $q$ and electronic state $m$ is described
by $\lambda_{mq}$. The time-dependent quantum transport through this model
Hamiltonian can be studied by the TDDFT-OS-NEGF method, it has been
shown that the EOMs automatically terminate at the second tier for
the non-interacting systems~\cite{yijing2008,zheng:114101}. 
However, in the presence of electron-phonon interaction, higher order tier EOMs emerge.
Previous attempt to investigate the
time-dependent quantum transport including electron-phonon interaction focuses
on weak coupling regime only. In this regime, lowest order expansion can be applied,
and the EOMs terminate at finite tier\cite{zhang:164121}.
However, the lowest order expansion approximation breaks down when the coupling strength becomes strong.
Hence it is desirable to go beyond the lowest order expansion. In this work,
a polaron transformation is applied to remove the
explicit electron-phonon coupling
term in the total Hamiltonian\cite{mahan2000book}: $\bar{H}=e^SHe^{-S}$. Since
\begin{equation}\label{eq:transform1}
e^S H e^{-S}= H+[S,H]+\frac{1}{2}[S,[S,H]]+\cdots
\end{equation}
and $e^S ABC e^{-S}=e^S A e^{-S}e^S B e^{-S}e^S C e^{-S}
=\bar{A} \bar{B}\bar{C}$, eliminating the 
explicit electron-phonon
coupling term requires $H_{ep}+[H_0,S]=0$ ($H_0=H_e+H_p$), it can be proven
that $S=\sum_{mq}\frac{\lambda_{mq}}{\omega_q}c^\dag_mc_m(a^\dag_q-a_q)$
satisfies the above condition, and the corresponding transformed Hamiltonian
reads
\begin{eqnarray}\label{eq:modelhamiltonian2}
\bar{H}
=&&\sum_m\bar{\epsilon}_mc^\dag_m c_m+\sum_{m\neq n}\bar{U}_{mn}c^\dag_m
c_mc^\dag_n c_n + \sum_{k_\alpha}
\epsilon_{k_\alpha}c^\dag_{k_\alpha}c_{k_\alpha}
\nonumber\\
&&+\sum_{mk_\alpha}[V_{k_\alpha m}c^\dag_{k_\alpha m}c_m X_m+
\text{H.c.}]
+\sum_q\omega_q a^\dag_a a_q.
\end{eqnarray}
where $X_m$ is the shift-operator, which is defined as
\begin{equation}
X_m=\text{exp}[-\sum_q\frac{\lambda_{mq}}{\omega_q}(a^\dag_q-a_q)].
\end{equation}
After the polaron transformation, there is no explicit electron-phonon
interaction term, phonon's influence on electrons is instead described by
three terms: (1) the polaron-shifted energies $\bar{\epsilon}_m
=\epsilon_m-\sum_q \frac{\lambda^2_{mq}}{\omega_q}$, which includes
the energy renormalization effect due to electron-phonon interaction;
(2) the phonon-mediated electron-electron interaction terms
$\bar{U}_{mn}=U_{mn}-2\sum_q\frac{\lambda_{mq}\lambda_{nq}}{\omega_q}$,
containing the effective electron-electron attractive interaction
mediated by phonon; (3)
coupling term between electronic states of device and lead, which
is renormalized by the shift operator $X_m$.
It is noted that strong electron-phonon interaction can result in a net
attractive interaction between electrons and consequently generates a
cooper pair in the superconductor.
As this article mainly focuses on the effect of electron-phonon coupling
on the electron transport properties, the effect of electron-electron
interaction is neglected by setting the renormalized electron-electron
interaction $\bar{U}$ to zero, i.e., choosing the original
electron-electron interaction strength $U$ to be the same as
$2\sum_q\frac{\lambda_{mq}\lambda_{nq}}{\omega_q}$.

\subsection{Time-dependent quantum transport theory with polaron 
transformation}
The key quantity in the NEGF method is the single-particle 
Green's function defined on the Keldysh contour, 
which is given by~\cite{C1CP21161G,nitzan2006prb,
PhysRevB.77.205314,PhysRevLett.107.046802}
\begin{eqnarray}\label{eq:nef}
G_{mn}(\tau,\tau')=&&-i\langle T_c c_m(\tau)c^\dag_n(\tau')\rangle_H
\nonumber\\
=&&-i\langle T_c c_m(\tau)X_m(\tau)c^\dag_n(\tau')X^\dag_n(\tau')\rangle_{\bar{H}},
\end{eqnarray}
where $\tau$ and $\tau'$ are the
time variables defined on the Keldysh contour, and $T_c$ is the contour time-ordering
operator. Eq.(\ref{eq:nef}) determines the dynamics of coupled electron 
and phonon, we employ the following approximation to decouple the electron 
and phonon dynamics~\cite{C1CP21161G,nitzan2006prb,
PhysRevB.77.205314,PhysRevLett.107.046802}
\begin{equation}\label{eq:nefdec}
G_{mn}(\tau,\tau')=\bar{G}_{mn}(\tau,\tau')K_{mn}(\tau,\tau'),
\end{equation}
where
\begin{eqnarray}
\bar{G}_{mn}(\tau,\tau')=&&-i\langle T_c c_m(\tau)c^\dag_n(\tau')\rangle_{\bar{H}}
\nonumber\\
K_{mn}(\tau,\tau')=&& \langle T_c X_m(\tau)X^\dag_n(\tau')\rangle_{\bar{H}}.
\end{eqnarray}
The decoupling in Eq.(\ref{eq:nefdec}) is inherent in the Born-Oppenheimer
approximation. Even the decoupling approximation is made, there is still
correlation between electron and phonon if self-consistent procedure is
operated \cite{nitzan2006prb}, which is similar to the diagram dressing
process in the standard many-body perturbation theory. In the following,
$\bar{G}(\tau,\tau')$ and $K(\tau,\tau')$ are referred as the electronic
Green's function and shift generator correlation function, respectively.

EOM of $\bar{G}(\tau,\tau')$ is very similar to
that of electronic Green's function of 
non-interacting system because
the transformed Hamiltonian $\bar{H}$ does not contain the explicit
electron-phonon interaction term. The only difference is that
the coupling term in $\bar{H}$ is different by a shift generator $X_m$.
If $X_m$ is replaced by its expectation
value $\langle X_m\rangle$, EOM of $\bar{G}$ reduces to the EOM of
non-interacting system, 
with $V_{k_\alpha m}$ replaced by $V_{k_\alpha m}\langle X_m\rangle\equiv\bar{V}_{k_\alpha m}$,
this method is regarded as the mean-field approach. Beyond
mean-field approach, employing the EOM of the electronic
Green's function $G_{mn}(\tau,\tau')$ gives
\begin{eqnarray}\label{eq:eomofgreenc}
i\partial_\tau \bar{G}_{mn}(\tau,\tau')=&&\delta(\tau-\tau')+
\sum_k h_{mk} \bar{G}_{kn}(\tau,\tau')
\nonumber\\&&
+\sum_{\alpha,k}\int d\tau_1 \Sigma_{\alpha,mk}(\tau,\tau_1)\bar{G}_{kn}(\tau_1,\tau')
\end{eqnarray}
where $h_{mk}\equiv\bar{\epsilon}_m\delta_{mk}$ and
the self-energy due to the coupling between device and lead $\alpha$
is given by
\begin{eqnarray}\label{eq:self1}
\Sigma_{\alpha,mn}(\tau,\tau')&&=\sum_{k_\alpha} V^*_{k_\alpha m}V_{k_\alpha n}
g_{k_\alpha}(\tau,\tau')\langle T_c X_n(\tau')X^\dag_m(\tau)\rangle_{\bar{H}}
\nonumber\\&&
\equiv \Sigma^0_{\alpha,mn}(\tau,\tau')K_{nm}(\tau',\tau),
\end{eqnarray}
where $g_{k_\alpha}(\tau,\tau')$ is the free Green's function for state $k_\alpha$
in the lead $\alpha$ defined on the Keldysh contour; $\Sigma^0_{\alpha}(\tau,\tau')$
is the self-energy without
electron-phonon coupling or within the untransformed Hamiltonian. 
Projecting Eq.(\ref{eq:eomofgreenc})
on real-time axis gives the EOM of 
the lesser component of Green's function $\bar{G}^<(t,t')$. Since
$\bar{\sigma}(t)=-i\bar{G}^<(t,t')|_{t=t'}$, EOM of density matrix with respect to
transformed Hamiltonian is
\begin{equation}\label{eq:eomofdm1}
i\partial_t \bar{\sigma}=[h,\bar{\sigma}]-\sum_\alpha [\varphi_\alpha(t)-\varphi^\dag_\alpha(t)],
\end{equation}
where
\begin{equation}\label{eq:varphialpha}
\varphi_\alpha(t)=i\int^t_{-\infty}dt_1 [
\bar{G}^<(t,t_1)\Sigma^>_\alpha(t_1,t)-
\bar{G}^>(t,t_1)\Sigma^<_\alpha(t_1,t)].
\end{equation}
Eqs.(\ref{eq:eomofdm1}) and (\ref{eq:varphialpha}) are similar to the
non-interacting case \cite{PhysRevB.87.085110}, the difference is that the
density matrix $\bar{\sigma}$ and self-energy are with
respect to the polaron transformed Hamiltonian and shift
generator correlation function is contained in the self-energy.

Aside from the electronic Green's function $\bar{G}(\tau,\tau')$, shift generator
correlation function $K(\tau,\tau')$ has also to be evaluated in order to obtain
the self-energy $\Sigma_\alpha(\tau,\tau')$.
Second-order cumulant expansion with respect to
the electron-phonon coupling strength $\lambda_{mq}$ leads to
\cite{nitzan2006prb,PhysRevB.87.085422}
\begin{eqnarray}
\langle T_c X_m(\tau)X^\dag_n(\tau')\rangle
=&&\text{exp}\Big[\sum_{qq'}i\frac{\lambda_{mq}\lambda_{nq'}}{\omega_q\omega_{q'}}
D_{qq'}(\tau,\tau')
\nonumber\\&&
-i\frac{\lambda_{mq}\lambda_{mq'}+\lambda_{nq}\lambda_{nq'}}{2\omega_q\omega_{q'}}
D_{qq'}(\tau,\tau)
\Big],
\end{eqnarray}
where the phonon Green's function is defined as
\begin{equation}
D_{qq'}(\tau,\tau')=-i\langle T_c P_q(\tau)P_{q'}(\tau')\rangle
\end{equation}
with momentum operator $P_q=-i(a_q-a^\dag_q)$.
Similar to the electronic Green's function, EOM of
$D_{qq'}(\tau,\tau')$ reads
\begin{eqnarray}\label{eq:eomphononc}
D^{0,-1}_q
D_{qq'}(\tau,\tau')=&&
\delta(\tau,\tau')+\sum_{q_1}\int d\tau
\nonumber\\&&\times \Pi_{qq_1}(\tau,\tau_1)D_{q_1q'}(\tau_1,\tau'),
\end{eqnarray}
The operator $D^{0,-1}_q$ in above equation is introduced as
$D^{0,-1}_q=-\frac{1}{2\omega_q}(\partial^2_\tau+\omega^2_q)$
with the property that
$
D^{0,-1}_q
D^0_{q}(\tau,\tau')=\delta(\tau,\tau'),
$
where $D^0_q(\tau,\tau')$ is the free phonon Green's function,
i.e., the Green's function decoupled from the electron.
$\Pi_{qq'}(\tau,\tau')$ in the Eq.(\ref{eq:eomphononc}) is the
corresponding self-energy to phonon Green's function accounting
for the electron-phonon interaction, its expression can be derived
in analogous to $\Sigma_\alpha(\tau,\tau')$:
\begin{eqnarray}\label{eq:selftophononc}
\Pi_{qq'}(\tau,\tau')=&&-i\sum_{mn}
\frac{\lambda_{mq}\lambda_{nq'}}{\omega_q\omega_{q'}}
\Big[
\Sigma_{mn}(\tau,\tau') \bar{G}_{nm}(\tau',\tau)
\nonumber\\&&+
\Sigma_{mn}(\tau',\tau) \bar{G}_{nm}(\tau,\tau')\Big].
\end{eqnarray}
Eqs.(\ref{eq:eomofgreenc})-(\ref{eq:selftophononc})
constitute a closed set of equations for electronic and phonon
Green's function of the non-equilibrium system with
strong electron-phonon interaction.
Since the self-energy to the phonon $\Pi_{qq'}(\tau,\tau')$
depends on the electronic Green's function $\bar{G}$ and the
shift-generator correlation function is included in the
self-energy to electron $\Sigma_\alpha$,
the EOMs of electronic and phonon Green's functions 
have to be be solved self-consistently.

\subsubsection{Observable of interest}
Transient current through lead $\alpha$
is determined by the number of electrons passing through
the interface between the lead $\alpha$ and device per unit
time,
\begin{eqnarray}
I_\alpha(t)=&&-\frac{d}{dt}\sum_{k_\alpha}\langle c^\dag_{k_\alpha}
c_{k_\alpha} \rangle_H
\nonumber\\
=&&2i\sum_{k_\alpha,m}\left[
V_{k_\alpha m}  \langle c^\dag_{k_\alpha} c_m X_m\rangle_{\bar{H}}-
\text{H.c.}
\right].
\end{eqnarray}
In terms of Green's function and self-energy, $I_\alpha(t)$
is expressed as
\begin{eqnarray}\label{eq:current}
I_\alpha(t)=&&
\int^t_{-\infty} d\tau \text{Tr}\Big[
\bar{G}^>(t,\tau)\Sigma^<_\alpha(\tau,t)
\nonumber\\&&
-\bar{G}^<(t,\tau)\Sigma^>_\alpha(\tau,t)
+\text{H.c.}
\Big]
\nonumber\\
=&&i\text{Tr}[\varphi_\alpha(t)-\varphi^\dag_\alpha(t)].
\end{eqnarray}
In the above equation, $\bar{G}^<$ and $\bar{G}^>$
are the lesser and greater Green's
functions of device, and $\Sigma^<_\alpha$
and $\Sigma^>_\alpha$
are the lesser and greater
self-energies due to the lead $\alpha$, respectively. The first term 
of Eq. (\ref{eq:current}) is interpreted as the out-coming rate of 
electron from device to lead $\alpha$ while the second term of
Eq. (\ref{eq:current}) is interpreted as the incoming rate of 
electron from lead a to device. Consequently, $\varphi_\alpha(t)$ 
corresponds to the net rate of electron going through the interface 
between lead $\alpha$ and device. Hence, transient current can be 
evaluated by the trace of the auxiliary density matrix.

\subsection{EOMs for auxiliary density matrices}
A closed set of EOMs has been established in the previous section.
Obviously, if the auxiliary density matrix $\varphi_\alpha(t)$
can be evaluated exactly, the density matrix can be obtained through
its EOM.

As described previously, shift generator
correlation function $K(\tau,\tau')$ is also required
to obtain the self-energy $\Sigma(\tau,\tau')$.
And $K(\tau,\tau')$ depends on the phonon
Green's function which is coupled with electronic Green's function
via its self-energy $\Pi_{qq'}(\tau,\tau')$.
Self-consistent calculation of phonon and electronic Green's function
is required in principle. However, numerical implementation
of self-consistent calculation for the transient regime is
non-trivial and computationally expensive. In practice,
we assume the phonon is in equilibrium and undressed by the electron.
The influence of electron to the phonon can be introduced
through a phenomenological rate equation including the renormalization,
damping and heating effect
\cite{PhysRevB.72.201101,PhysRevLett.93.256601,Kristen2013}.
With the assumption that phonon is in the
equilibrium and undressed by electron, the shift generator correlation
function can be rewritten in a simple form \cite{mahan2000book,
PhysRevB.71.165324,nitzan2006prb}.
The lesser projection of shift generator correlation function is
expressed as
\begin{eqnarray}\label{eq:shift1}
K^<_{mn}(t,t')=&&\langle X^\dag_n(t')X_m(t)\rangle.
\nonumber\\
=&&\prod^M_{q=1} \Bigg\{
e^{-\frac{\lambda^2_{mq}+\lambda^2_{nq}}{2\omega^2_q}(2N_q+1)}
\text{exp}
\Big\{\frac{\lambda_{mq}\lambda_{nq}}{\omega^2_q}
\times\nonumber\\
&&\Big[
N_q e^{-i\omega_q(t-t')}
+(N_q+1)e^{i\omega_q(t-t')}
\Big]\Big\}\Bigg\},
\end{eqnarray}
where $N_q$ is the occupation number for the $q$th phonon mode
determined by Bose-Einstein distribution function,
$M$ is the number of phonon modes.
The lesser $K^<_{mn}(t,t')$ can be decomposed as
\begin{eqnarray}\label{eq:shift2}
K^<_{mn}(t,t')
=&&\prod^M_{q=1}
\left[ \sum_{p_q}
L^{p_q}_{mn} e^{i p_q\omega_q(t-t')}
\right]
\nonumber\\
=&&\sum_{p_1 p_2\cdots p_M}
L^{p_1}_{mn}L^{p_2}_{mn}\cdots
L^{p_M}_{mn}e^{i\bm{p}^\text{T} \bm{\omega}(t-t')}
\nonumber\\
\equiv&&
\sum_{\bm{p}}L^{\bm{p}}_{mn}e^{i\bm{p}^\text{T} \bm{\omega}(t-t')},
\end{eqnarray}
where both $\bm{p}$ and $\bm{\omega}$
are row vectors,
$\bm{p}^\text{T}\bm{\omega}=\sum_q p_q\omega_q$.
And $L^{\bm{p}}_{mn}=L^{p_1}_{mn}L^{p_2}_{mn}\cdots
L^{p_M}_{mn}$, where $L^{p_q}_{mn}$ is the modified Bessel function
\begin{eqnarray}
L^{p_q}_{mn}=&&e^{-\frac{\lambda^2_{mq}+\lambda^2_{nq}}{2\omega^2_q}(2N_q+1)}
e^{p_q\omega_q\beta/2}
\nonumber\\&&\times 
I_{p_q}\left(\frac{2\lambda_{mq}\lambda_{nq}}{\omega^2_q}\sqrt{N_q(N_q+1)}\right),
\end{eqnarray}
$I_{p_q}$ is the $p_q$th order Bessel function.
From the expression of $K^<_{mn}(t,t')$, it is obvious that
\[
K^{<}_{mn}(t,t)=\sum_{\bm{p}}L^{\bm{p}}_{mn}=\prod^M_{q=1}
e^{-\frac{(\lambda_{mq}-\lambda_{nq})^2}{2\omega^2_q}(2N_q+1)}.
\]
The greater projection of shift generator correlation function is
$
K^>_{mn}(t,t')=\langle X_m(t)X^\dag_n(t')\rangle
=[K^<_{mn}(t,t')]^\dag.
$
It can be verified that $K^<(t,t')\simeq K^>(t,t')$ in the
high-temperature limit where $N_q\simeq N_q +1$.
This can be regarded as neglecting the Fermi sea
\cite{huanghartmut2008,mahan2000book,PhysRevB.66.075303,
PhysRevB.50.5528,PhysRevB.67.165326,PhysRevB.40.11834}.

The mean-field approach to the shift generator $X_i$ in the device-lead
coupling term leads to a simple form of self-energies:
\begin{equation}\label{eq:selfm}
\Sigma_{\alpha,mn}(\tau,\tau')=\sum_{k_\alpha} \bar{V}^*_{mk_\alpha}
\bar{V}_{k_\alpha n}g_{k_\alpha}(\tau,\tau').
\end{equation}
Obviously, Eqs.(\ref{eq:eomofgreenc}) and (\ref{eq:selfm}) are same
as the non-interacting case \cite{PhysRevB.50.5528}.
Hence, with the mean-field approximation to $V_{k_\alpha m}X_m$, the EOMs of
density matrix $\bar{\sigma}$ and auxiliary density matrices $\varphi_\alpha(t)$
are the same as the non-interacting case, and the method to evaluate
the time-dependent density matrix $\bar{\sigma}$ and auxiliary density
matrices $\varphi_\alpha(t)$ has been developed previously
\cite{PhysRevB.87.085110}.

Without the mean-field approximation to  $V_{k_\alpha m}X_m$,
the self-energy is described by Eq.(\ref{eq:self1})
which contains the shift-generator correlation function.
The inclusion of shift-generator correlation function in the
self-energies makes the evaluation of auxiliary density matrices more
complicated. The lesser (greater) self-energy in
Eq.(\ref{eq:varphialpha}) can be obtained by projecting
Eq.(\ref{eq:self1}) on real-time axis:
\begin{equation}\label{eq:self2}
\Sigma^{\gtrless}_{\alpha,mn}(t',t)=
\Sigma^{0,\gtrless}_{\alpha,mn}(t',t)K^{\lessgtr}_{nm}(t,t').
\end{equation}
For the $\Sigma^{0,\gtrless}_\alpha(t',t)$, we have shown previously
that it can be decomposed into series according to the Pad\'{e}
expansion of Fermi function \cite{PhysRevB.87.085110}. In particular,
WBL approximation leads to a simple form
of time-dependent lesser (greater) self-energy:
\begin{equation}
\Sigma^{0,\gtrless}_{\alpha}(t',t)=\mp \frac{i}{2}\delta(t-t')\Lambda^0_\alpha
+\text{sgn}(t-t')\sum^N_k \Sigma^{\text{sgn}(t-t')}_{\bm{a}}(t',t),
\end{equation}
where the sign of first term is $-$ ($+$) for the greater (lesser)
self-energy, $\text{sgn}(t-t')$ is the sign function and
$\Lambda^0_\alpha=\pi\sum_{k_\alpha}|V|^2
\delta(\epsilon_f-\epsilon_{k_\alpha})$
is the line-width function evaluated at the Fermi energy.
$\Sigma^{\text{sgn}(t-t')}_{\bm{a}}(t',t)$ is the component of
self-energy due to the Pad\'{e} expansion, which is defined as
(a notation $\bm{a}=\alpha k$ is used)
\begin{equation}
\Sigma^{\text{sgn}(t-t')}_{\bm{a}}(t',t)=\frac{2}{\beta}\eta_k
e^{i\int^t_{t'} \epsilon^{\text{sgn}(t-t')}_{\bm{a}}(t_1)dt_1}
\Lambda^0_\alpha.
\end{equation}
Here $\epsilon^\pm_{\bm{a}}(t)=\pm i\zeta_k/\beta+\mu_\alpha+\Delta_\alpha(t)$.
The $\pm i\zeta_k/\beta+\mu_\alpha$ are the $k$th Pad\'{e} poles in the
upper and lower half plane, respectively. $\eta_k/\beta$ is the
corresponding coefficient. $\beta$ is the inverse temperature and
$\Delta_\alpha(t)$ is the applied time-dependent bias voltage.
Based on the approximation to the bare self-energy
$\Sigma^{0,\gtrless}_\alpha(t',t)$,
the lesser (greater) self-energy can be rewritten as
\begin{equation}
\Sigma^{\gtrless}_{\alpha}(t',t)=\mp \frac{i}{2}\delta(t-t')\Lambda_\alpha
+\text{sgn}(t-t')\sum^N_k
\Sigma^{\lessgtr,\text{sgn}(t-t')}_{\bm{a}}(t',t),
\end{equation}
where
\[
\Lambda_{\alpha,mn}=\Lambda^0_{\alpha,mn}K^<_{nm}(t,t),
\]
and
\[
\Sigma^{\lessgtr,\text{sgn}(t-t')}_{\bm{a},mn}(\tau,t)=
\Sigma^{\text{sgn}(t-t')}_{\bm{a},mn}(\tau,t)
K^{\lessgtr}_{nm}(t,\tau).
\]
As a result, the auxiliary density matrix
$\varphi_\alpha(t)$ is rewritten as
\begin{eqnarray}\label{eq:varphialpha2}
\varphi_\alpha(t)=&&i[\sigma(t)-1/2]\Lambda_\alpha+
\sum^N_k \varphi_{\bm{a}}(t).
\end{eqnarray}
The first term on the right-hand side (RHS) of above
equation comes from the integration over lesser/greater Green's
function $G^{\gtrless}(t,\tau)$ and the delta function $\delta(t-\tau)$;
The second term on the RHS of Eq.(\ref{eq:varphialpha2}) is
\begin{eqnarray}\label{eq:varphialphak}
\varphi_{\bm{a}}(t)=&&
i\int^t_{-\infty}d\tau \bar{G}^>(t,\tau)
[\Sigma^{<,+}_{\bm{a}}(\tau,t)-\Sigma^{>,+}_{\bm{a}}(\tau,t)]
\nonumber\\&&-
i\int^t_{-\infty} d\tau \bar{G}^r(t,\tau)\Sigma^{<,+}_{\bm{a}}(\tau,t).
\end{eqnarray}
The $\varphi_{\bm{a}}(t)$ is the component of the first tier
auxiliary density matrix, the number of which is determined by
the order of Pad\'{e} expansion.

With the Pad\'{e} approximation to Fermi function
and WBL approximation to self-energy, the time-dependent
quantum transport problem with strong electron-phonon interaction
can be solved through the EOM of $\bar{\sigma}(t)$ once
$\varphi_{\bm{a}}(t)$ is known. The difficulty of evaluation
of $\varphi_{\bm{a}}(t)$ lies in the lesser (greater)
shift-generator correlation function $K^{\gtrless}_{nm}(t,\tau)$.
In absence of electron-phonon coupling, the shift-generator correlation
function $K(t,\tau)=1$, then
$\Sigma^{\gtrless,+}_{\alpha k}=\Sigma^{+}_{\alpha k}$
and $\varphi_{\bm{a}}(t)$ reduces to
\begin{equation}\label{eq:varphialpha0}
\varphi_{\bm{a}}(t)=-i\int^\infty_{-\infty}
\bar{G}^r(t,\tau)\Sigma^{+}_{\bm{a}}(\tau,t)d\tau,
\end{equation}
which can be solved through its EOM since EOM of $G^r(t,\tau)$ is
closed under WBL approximation \cite{PhysRevB.87.085110}.
In contrast, in presence of
electron-phonon coupling, $\varphi_{\bm{a}}(t)$
does not have the simple form as Eq.(\ref{eq:varphialpha0})
due to the difference between $K^<(t,t')$ and $K^>(t,t')$.
In order to obtain the solution to $\varphi_{\bm{a}}(t)$
in presence of electron-phonon interaction,
an efficient method of evaluation of $\varphi_{\bm{a}}(t)$
has to be developed.

According to the expansion of $K^{\gtrless}_{nm}(t,\tau)$
in Eq.(\ref{eq:shift2}),
decomposition can be further applied to $\varphi_{\bm{a}}(t)$
and each component can be solved through its EOM.
As shown previously, the difference
between $K^>(t,t')$ and $K^<(t,t')$ becomes smaller with increasing
temperature since $N_q\simeq N_q+1$ at high temperature. Hence,
we will discuss the solution to $\varphi_{\bm{a}}(t)$ in two different
regimes, i.e., high and low temperature regimes.

\subsubsection{High-temperature limit}
At high phonon temperature, i.e., $N_q\gg 1$ and
$N_q\simeq N_q+1$, it is easy to verify that the lesser and greater
projection of shift-generator correlation have the relation
$K^>_{mn}(t,t')\simeq K^<_{nm}(t,t')$ in
the high-temperature limit, therefore
$\Sigma^{>,+}_{\bm{a}}(\tau,t)=\Sigma^{<,+}_{\bm{a}}(\tau,t)$
and the first term on the RHS of Eq.(\ref{eq:varphialphak}) vanishes.
Based on the expansion of shift-generator correlation function
described by Eq.(\ref{eq:shift2}),
$\Sigma^{<,+}_{\bm{a}}(\tau,t)$ can be decomposed as
\[
\Sigma^{\gtrless,+}_{\bm{a}}(\tau,t)=\sum_{\bm{p}}
\Sigma^{+}_{\bm{ap}}(\tau,t).
\]
where
\[
\Sigma^{+}_{\bm{ap},mn}(\tau,t)=
\Sigma^+_{\bm{a},mn}(\tau,t) L^{\bm{p}}_{nm}
e^{i\bm{p}^{\text{T}}\bm{\omega}(t-\tau)}.
\]
Accordingly, $\varphi_{\bm{p}}(t)$ can be further decomposed
into $\sum_{\bm{p}} \varphi_{\bm{ap}}(t)$ where
\begin{equation}\label{eq:varphinalphak}
\varphi_{\bm{ap}}(t)=-i\int^\infty_{-\infty}d\tau
\bar{G}^r(t,\tau)\Sigma^{+}_{\bm{ap}}(\tau,t).
\end{equation}
The definition of $\varphi_{\bm{ap}}(t)$ is similar to
Eq.(\ref{eq:varphialpha0}) except the self-energy is replaced
by the phonon dressed one. Analogous to the non-interacting case,
EOM of $\varphi_{\bm{ap}}(t)$ is self-closed since
EOMs of $\bar{G}^r(t,\tau)$ and $\Sigma^{+}_{\bm{ap}}(\tau,t)$
are both self-closed, i.e.,
\begin{equation}\label{eq:eomvarphinalphak}
i\dot{\varphi}_{\bm{ap}}(t)
=-i\frac{2\eta_k}{\beta}\tilde{\Lambda}^{\bm{p}}_\alpha
-[\epsilon^+_{\bm{a}}(t)+\bm{p}^{\text{T}}\bm{\omega}-h(t)+i\Lambda]
\varphi_{\bm{ap}}(t),
\end{equation}
where $\tilde{\Lambda}^{\bm{p}}_{\alpha,mn}=\Lambda^0_{\alpha,mn}L^{\bm{p}}_{nm}$.
Hence, just like in absence of electron-phonon interaction, the
TDDFT-OS-NEGF-WBL terminates at the first tier in the high-temperature limit.
Solutions to the density matrix and auxiliary ones can be readily evaluated
through their EOMs with corresponding initial conditions.

\subsubsection{Low temperature}
At low temperature, the relation
$K^>_{mn}(t,t')\simeq K^<_{mn}(t,t')$ is not valid
since $N_q\simeq N_q+1$ no longer holds, especially
$N_q$ vanishes at the zero temperature limit.
Hence, the first term of Eq.(\ref{eq:varphialphak}) does not vanish and
$\varphi_{\bm{a}}(t)$ cannot be decomposed into the simple
form as Eq.(\ref{eq:varphinalphak})
due to the difference between $K^<(t,\tau)$ and $K^>(t,\tau)$
at low temperature. Though the difference between $K^<(t,\tau)$ and
$K^>(t,\tau)$, $\Sigma^{\gtrless,+}_{\bm{a}}(\tau,t)$ can be
decomposed separately as
\[
\Sigma^{\gtrless,+}_{\bm{a}}(\tau,t)=\sum_{\bm{p}}
\Sigma^{\gtrless,+}_{\bm{ap}}(\tau,t).
\]
where
\[
\Sigma^{\gtrless,+}_{\bm{ap},mn}(\tau,t)=
\Sigma^+_{\bm{a},mn}(\tau,t) L^{\bm{p}}_{nm}
e^{\mp i\bm{p}^\text{T}\bm{\omega}(t-\tau)}.
\]
As a consequence, Eq.(\ref{eq:varphialphak}) is rewritten as
\begin{eqnarray}\label{eq:varphialphak2}
\varphi_{\bm{a}}(t)=\sum_{\bm{p}} [\varphi_{\bm{ap}}(t)+
\varphi^<_{\bm{ap}}(t)-\varphi^>_{\bm{ap}}(t)],
\end{eqnarray}
where the definition of $\varphi_{\bm{ap}}(t)$ is same
as Eq.(\ref{eq:varphinalphak}) and $\varphi^{\gtrless}_{\bm{ap}}(t)$
is given by
\begin{equation}\label{eq:phiakgtrless}
\varphi^{\gtrless}_{\bm{ap}}(t)=
i\int^t_{-\infty} d\tau \bar{G}^{>}(t,\tau)
\Sigma^{\gtrless,+}_{\bm{ap}}(\tau,t).
\end{equation}

Both $\varphi_{\bm{ap}}(t)$ and $\varphi^{\gtrless}_{\bm{ap}}(t)$
can be solved by their EOMs. The EOM of $\varphi_{\bm{ap}}(t)$ is
same as Eq.(\ref{eq:eomvarphinalphak}).
Since EOMs of $\bar{G}^{>}(t,\tau)$ is not closed,
higher tier components appear in the EOMs
of $\varphi^{\gtrless}_{\bm{ap}}(t)$.
EOM of $\varphi^{\gtrless}_{\bm{ap}}(t)$ is written as
\begin{eqnarray}
i\dot{\varphi}^{\gtrless}_{\bm{ap}}(t)=&&
-[\epsilon^+_{\bm{a}}(t)\mp
\bm{p}^\text{T}\bm{\omega}-h(t)+i\Lambda]
\varphi^{\gtrless}_{\bm{ap}}(t)
\nonumber\\&&
-i[\sigma(t)-1]
\frac{2\eta_k}{\beta}\tilde{\Lambda}^{\bm{p}}_\alpha
+\sum_{\bm{a}'}
\varphi^{\gtrless}_{\bm{a}',\bm{ap}}(t)
\end{eqnarray}
In above equation, $-$($+$) stands for the greater(lesser) component.
$\varphi^{\gtrless}_{\bm{a}',\bm{ap}}(t)$ is defined as
\begin{eqnarray}
\varphi^{\gtrless}_{\bm{a}',\bm{ap}}(t)=&&
-i\int^t_{-\infty}dt_1\int^t_{-\infty}d\tau
\Sigma^{-}_{\bm{a}'}(t,t_1)\bar{G}^a(t_1,\tau)
\Sigma^{\gtrless,+}_{\bm{ap}}(\tau,t)
\nonumber
\end{eqnarray}

It is obvious that the EOMs of $\varphi^{\gtrless}_{\bm{a}',\bm{ap}}(t)$
are closed, which is
\begin{eqnarray}
i\dot{\varphi}^{\gtrless}_{\bm{a}',\bm{ap}}(t)=&&
[\epsilon^{-}_{\bm{a}'} -\epsilon^+_{\bm{a}}\pm
\bm{p}^\text{T}\bm{\omega}]
\varphi^{\gtrless}_{\bm{a}',\bm{ap}}(t)
\nonumber\\&&
-i[\varphi_{\bm{a}'}(t)]^\dag \frac{2}{\beta}\eta_k
\tilde{\Lambda}^{\bm{p}}_\alpha
\end{eqnarray}
Thus, we get a closed set of EOMs for the electron transport with
electron-phonon interaction in the low temperature regime.
Compared to the high-temperature limit, an additional tier
appears as a result of the difference between lesser and
greater shift-generator correlation function.

\section{\label{sec:results}Results}
\subsection{Quantum interference in absence of electron-phonon 
interaction}
In this section, the methodology developed in the previous
section is used to study the quantum interference effects
in real-time dynamics of molecular junctions.

For simplification, quasi-degenerate two-state model systems 
are studied. The systems are coupled to two leads with
different chemical potential, where the electrons in one lead with
higher chemical potential can transfer 
via the system to the other lead. The two states in the systems
may couple differently to the leads, hence the electrons transfer 
from one lead through different states may end up with different 
phase when arriving at
another lead. The phase difference can induce constructive or
destructive interference effect in the electron transport.
The system setups and related parameters are summarized in the 
Table.~\ref{tab:models}. Even though the two-state
model simplifies the problem of quantum interference and phonon-induced decoherence, 
it captures the fundamental mechanism. The parameters of the model
can be fitted from the first-principles calculations and the model has been employed
to explain the experimental observation~\cite{PhysRevLett.109.056801}.
\begin{table}
\caption{\label{tab:models}
Parameters for the Models in the unit of eV.
}
\begin{ruledtabular}
\begin{tabular}{cccccrccc}
\textrm{Model}&
\textrm{$\epsilon_1$}&
\textrm{$\epsilon_2$}&
\textrm{$V_{L,1/2}$} &
\textrm{$V_{R,1}$} &
\textrm{$V_{R,2}$} &
\textrm{$\omega$}  &
\textrm{$\lambda_1$} &
\textrm{$\lambda_2$} \\
\colrule
 A & -0.005 & 0.005 & $v$ & $v$ & $v$  & $$    & $$  & $$    \\
 B & -0.005 & 0.005 & $v$ & $v$ & $-v$ & $$    & $$  & $$    \\
 C & -0.005 & 0.005 & $v$ & $v$ & $-v$ & $0.1$ & $0$ & $\lambda$
\end{tabular}
\end{ruledtabular}
\end{table}

The model A and model B have same parameters except one of the 
coupling constants has different sign. 
The different sign reflects 
the  different spatial symmetry of the two states, which represents
symmetric and antisymmetric combinations of localized molecular
orbital\cite{PhysRevB.87.085422}. Both models A and B have been
extensively studied before\cite{PhysRevB.87.085422}, which are well-suited
for the investigation of phonon-induced decoherence dynamics.
In this study, the coupling between system and leads is set
to be $v=0.5$~eV, and the hopping parameter in the leads
is $t=2$~eV. Given the hopping parameters in the leads,
the line-width function at Fermi energy is given by
Eq.(\ref{eq:linewidth}) as $|\Gamma_{\alpha,mn}|=2v^2/t=0.25$~eV.
Thus the leads induced broadening of the two states are $\sim 0.5$~eV,
corresponding to a life-time of $\sim 1.3$~fs.

\begin{figure}[!htb]
\includegraphics[width=0.48\textwidth]{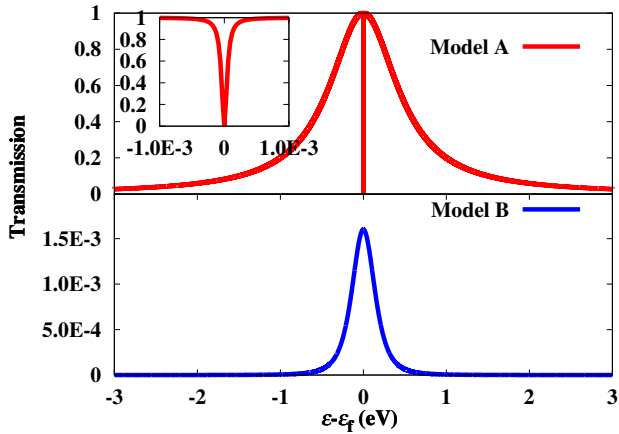}
\caption{\label{fig:trans0}
Transmission coefficient of models A and B. Comparison between
the transmissions of the two models indicates the strong
suppression of transmission due to the destructive
quantum interference. The inset 
shows the transmission coefficient of model A near the
Fermi level, which has very low transmission in this regime
due to antiresonance.}
\end{figure}

Figure.~\ref{fig:trans0} shows the transmission coefficient of the two models.
Due to the destructive quantum interference effect, the transmission
coefficient of model B is suppressed by at least 3 order of magnitude compared
to that of model A. The suppression of transmission due to quantum
interference effect in model B applies throughout the whole energy
range, including resonant and off-resonant regime. The presence of
destructive quantum interference effect in model B comes from
the outgoing wave function associated with the tunneling process
through state 2 to the right lead, which has $\pi$ phase difference
from the tunneling through state 1 to the right lead. This
phase difference arises from the different spatial structure of
the two states and is indicated by the different sign of 
system-lead coupling strengths to right
lead \cite{PhysRevB.87.085422}. In contrast to model B,
model A differs from it by the sign of coupling strength
between state 2 and right lead, i.e., $V_{R,2}$. As
shown in Fig.~\ref{fig:trans0}, transmission of model A
does not show destructive quantum interference
effect except in a narrow range in $[\epsilon_1:\epsilon_2]$.
The low transmission in this range is due to the
antiresonance \cite{nl901554s}.

\begin{figure}[!htb]
  \includegraphics[width=0.48\textwidth]{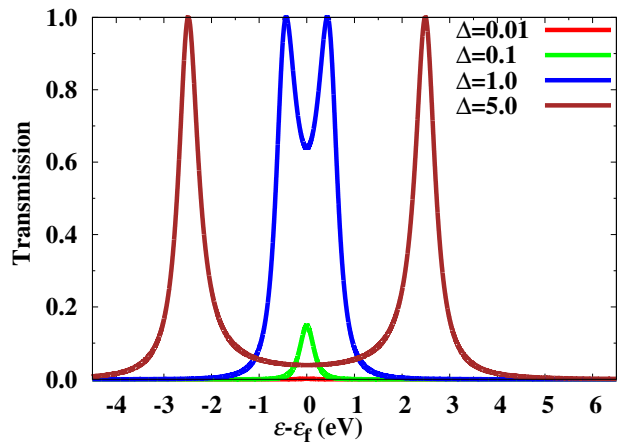}
  \caption{\label{fig:transcompare}Transmission of model B
  with different energy gap ($\Delta$).} 
\end{figure}

The energy gap between the two states 
($\Delta=\epsilon_2-\epsilon_1$) is designed to be small compared
to the line-width, i.e. $\Delta < \Gamma$, where 
$\Gamma=\sum_\alpha \Gamma_\alpha$.
This is very similar to the optical interference of double-slit 
which requires the width of double-slit to be small compared to the
light wavelength. If $\Delta\gg\Gamma$, 
electrons transport through the two states independently,
quantum interference cannot be observed. Fig.~\ref{fig:transcompare}
plots the transmission of model B with different energy gap.
The transmission increases with increasing energy gap.
At last, the transmission shows two conduction channels
when $\Delta\gg\Gamma$, which indicates that quantum 
interference dims out with increasing energy gap.

\begin{figure}
\includegraphics[width=0.45\textwidth]{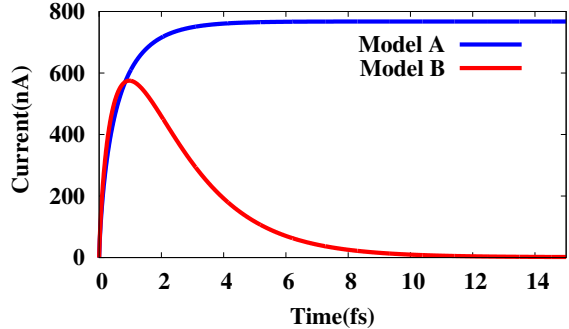}
\caption{\label{fig:curr_non}
Transient currents of model A and B. The real-time
dynamic of model B demonstrates the formation of destructive
quantum interference effect, resulting in a very low
state current when the interference is established.
The time-dependent
bias voltage is exponential growth type,
$V_L(t)=-V_R(t)=V_0(1-e^{-t/a})$ where
$V_0=5$~meV and $a=0.1$~fs.}
\end{figure}

Next, the dynamics of the two models under time-dependent bias voltage
are examined. The systems are in equilibrium state
before turning on bias voltage. After the time-dependent bias voltages
are applied on the leads, the systems are driven out of equilibrium.
In this study, the applied bias voltage is applied in a symmetric way:
$V_L(t)=-V_R(t)=V_0(1-e^{-t/a})$ where $V_0=5$~meV and $a=0.1$~fs.
The voltage is designed to be turned on quickly, i.e., $a<\hbar/\Gamma$,
in order to ensure the time-scale of the dynamics is dominated by the intrinsic
time-scale of the system itself. The transient currents of models A and
B are represented in Fig.~\ref{fig:curr_non}. As mentioned before,
the life-time of the states is $\sim 1.3$~fs, transient current
quickly reach its steady state in several femto-seconds.
Compared to model A, transient current of 
model B shows similar behaviour at the very beginning after the turning 
on the bias voltage. It begins to deviate from model A after about 
1~fs where quantum interference begins to take effect as electrons with $\pi$ phase difference reach the right lead.
The $\pi$ phase difference is accumulated in the transport process 
from the two states to the right lead
with increasing number of electrons reaching the right
lead via the two states, resulting in more pronounced destructive quantum interference
effect. As a consequence, the current of model B
diminishes as approaching steady state.

Due to the destructive quantum interference effects, the transient
current of model B reaches a rather low value ($\sim 1.2$~nA)
in the steady state. It is about
3 order of magnitude smaller compared to the steady
state current of model A, which is consistent
with the difference between the transmissions of the two models 
shown in Fig.~\ref{fig:trans0}.
The evolution of the transient current indicated by Fig.~\ref{fig:curr_non}
clearly demonstrates the transient formation of quantum interference effect
in the junctions, which requires a finite time to establish.

\subsection{Dynamics of decoherence in presence of electron-phonon 
interaction}
In the realistic devices, electron has the possibility of losing phase
coherence through the scattering by phonon or other phase-breaking 
mechanism. In this section, the effects of electron-phonon coupling
on quantum interference phenomena, i.e., decoherence dynamics, 
is examined. In this study, only a single vibrational mode
is considered for simplicity. Besides, the phonon is 
assumed to only coupled to one of the states.
The corresponding model is listed in the
Table.~\ref{tab:models} as model C.
\begin{figure}
\includegraphics[width=0.238\textwidth]{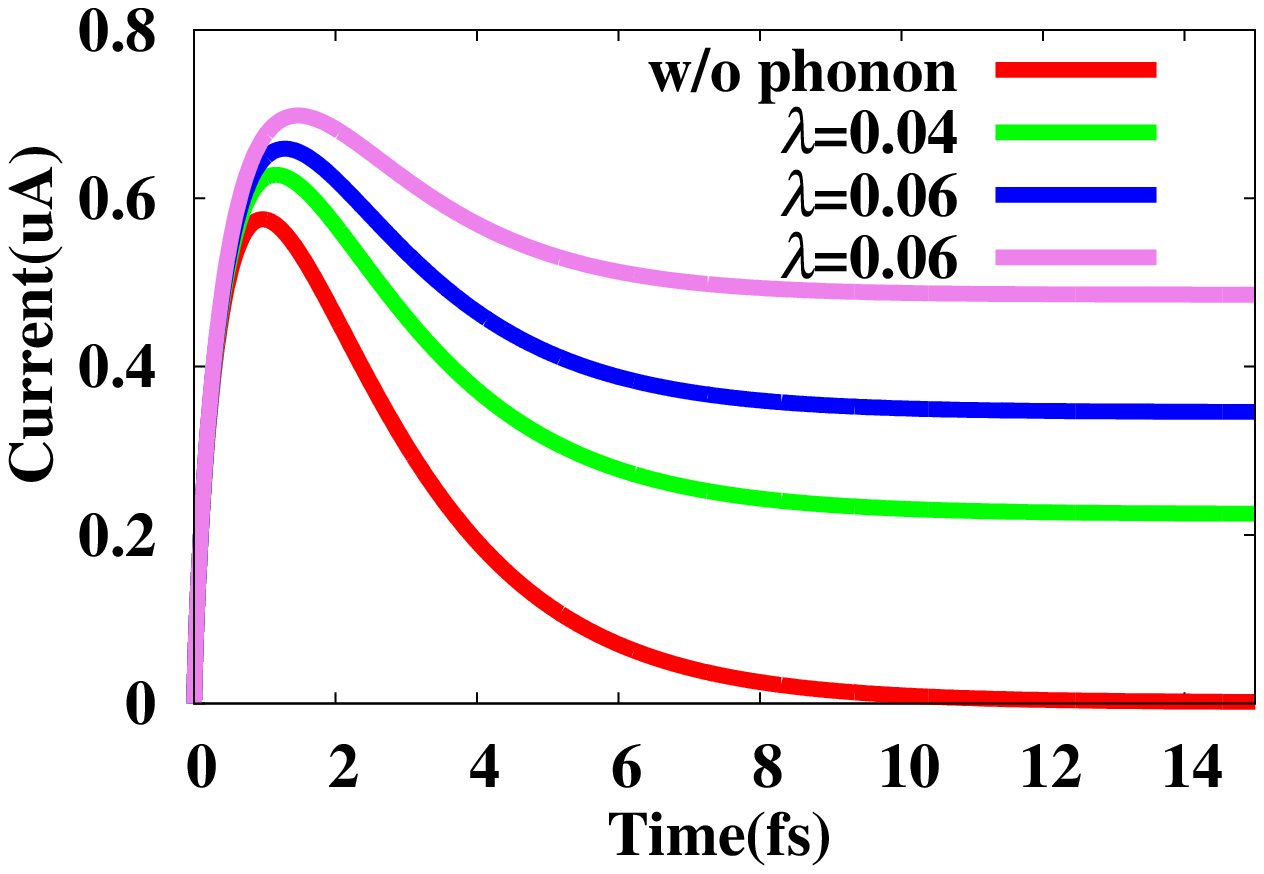}
\includegraphics[width=0.238\textwidth]{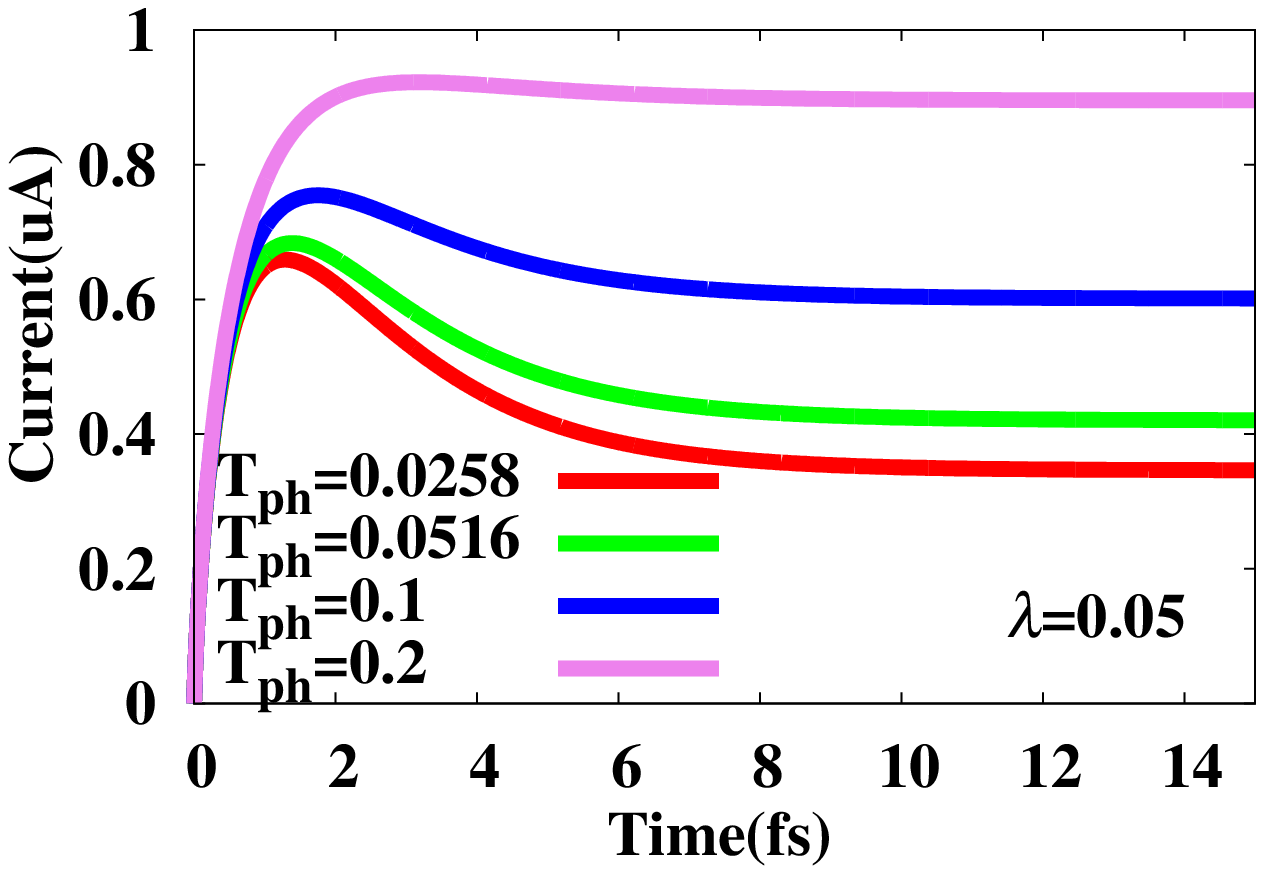}
\caption{\label{fig:curr_ph}
Transient currents of model C with different
setups.
Left panel: transient currents of four setups with
different electron-phonon coupling constant;
Right panel: transient currents of four setups with
different phononic temperature.
The time-dependent bias voltage is
$V_L(t)=-V_R(t)=V_0(1-e^{-t/a})$, where
$V_0=5$~meV and $a=0.1$~fs.}
\end{figure}

The quantum dynamics of model C are studied by applying
a time-dependent bias voltage which is same as the non-interacting
case. The system is initially in equilibrium state with 
electron-phonon interaction before turning on the bias voltage.
Fig.~\ref{fig:curr_ph} shows the transient currents
of model C with different setups. On the left panel, transient currents
with different electron-phonon coupling strength are
demonstrated, the temperatures of leads and phonon
are all set to be $T=0.0258$~eV, corresponding to
the room temperature. It clearly shows that the introduction
of electron-phonon interaction pronouncedly increases
the steady state current, which is due to the decoherence
in presence of phonon. Shortly after turning on the
bias voltage, the four different setups show similar dynamics,
this regime is related to the tunneling event of electrons from
left lead to the two states.
When electrons start to tunnel from
the two states to right lead, transient
currents begin to deviate. The reason
is that the inelastic scattering by phonon partly destroys 
the phase coherence between the transport electrons.
Stronger the electron-phonon coupling is,
more electrons will be scattered by phonons and the coherence
is further destroyed. As a result, the interference between
the tunneling electrons from the two different states
is significantly suppressed by phonon scattering, and steady 
state current shows monotonous relation with electron-phonon coupling
strength as indicated in the left panel of Fig.~\ref{fig:curr_ph}.

The right panel of Fig.~\ref{fig:curr_ph} shows the
transient currents with different phononic temperature.
The electron-phonon coupling constant is set as $\lambda=0.05$~eV.
At higher temperature, more phonons are occupied and hence the probability of
electrons being scattered is increased, and accordingly suppresses the
quantum interference effect. Consequently, the current
is enhanced by increasing the phononic temperature as
indicated in the right panel of Fig.~\ref{fig:curr_ph}.

\section{\label{sec:conclusion}Summary}
In this work, a dissipative time-dependent quantum transport
theory in the strong electron-phonon interaction regime is established through
the combination of TDDFT-OS-NEGF-WBL method and polaron transformation.
The polaron transformation avoids the explicit electron-phonon
coupling term, the effect of phonon on electron is transformed
to the polaron shifted energies, phonon mediated effective electron-electron
interaction and dressed device-lead coupling. 
In the high temperature limit,
neglecting the difference between lesser and greater
shift-generator correlation function results in a simple
EOM formalism which terminates at the first tier similarly
as the non interacting case. For the low temperature, second
tier auxiliary density matrix arises, and corresponding
EOMs are derived. It is worth noted that we demonstrate in this work the validity
of TDDFT-OS with tight-binding model for simplicity. Since the formalism established
in this work is based on single-particle theory, it can be readily implemented with TDDFT.
Within TDDFT, the correlation-effect
which is neglected in this work can also be taken into account through
exchange-correlation functional.

The dissipative time-dependent quantum transport theory in
the strong electron-phonon interaction regime is
applied to study the quantum interference and phonon-induced
decoherence in the molecular junctions. In absence of electron-phonon
interaction, transient current of the interference model system
clearly shows the transient effect of quantum interference, i.e. it undergoes
a nonequilibrium process before the interference pattern is formed.
The interference effect is reflected in the transient current of the system.
Shortly after the turning
on the bias voltage, the transient current increases before the interference effect is established.
As the destructive interference pattern forms when electrons reach the right lead, it suppresses significantly the current.
As a result, the transient current of
quantum interference system presents an overshot in the initial trace and diminishes in the long time limit.
The introduction of electron-phonon interaction scatters electrons
when they transport through
the junction. The scattering process breaks the phase coherence between
electrons from the two different states. This decoherence effect
due to electron-phonon scattering breaks the quantum
interference effect, resulting in pronounced increase of current.

\begin{acknowledgments}
The support from the Hong Kong Research Grant
Council (Contract Nos. HKU 7009/12P, 7007/11P
and 700913P (GHC)), the University Grant Council
(Contract No. AoE/P-04/08 (GHC)), National Natural Science 
Foundation of China (NSFC 21322306 (CYY),
NSFC 21273186 (GHC, CYY)) and National Basic Research Program of China (2014CB921402 (CYY)) is gratefully acknowledged. 
\end{acknowledgments}

\bibliography{ref}

\end{document}